\documentclass[french]{pfia}
\usepackage{amsmath}
\usepackage{graphicx}
\usepackage{array}
\newcolumntype{P}[1]{>{\centering\arraybackslash}p{#1}}
\usepackage{listings}
\usepackage{multirow}
\usepackage{xcolor}
\usepackage{listingsutf8}
\usepackage{soul}

\newcommand*{\I}{\imath}
\title{\textbf{Apport des ontologies pour le calcul de la similarité sémantique au sein d'un système de recommandation}}
\author{LE Ngoc Luyen\fup{1,2}, Marie-Hélène ABEL\fup{1}, Philippe GOUSPILLOU\fup{2}\\[6pt]
	\fup{1} Université de technologie de Compiègne, CNRS, Heudiasyc (Heuristics
	\\  and Diagnosis of Complex Systems), CS 60319 - 60203 Compiègne Cedex, France\\
	\fup{2} Vivocaz, 8 B Rue de la Gare, 002200, Mercin-et-Vaux, France\\
}

\date{}

\begin{document}
	
	\maketitle
	
	% ------------------------------------------
	% RÉSUMÉS ET MOTS-CLÉS
	% ------------------------------------------
	
	\begin{resume}
		La mesure de la parenté ou ressemblance sémantique  entre les termes, les mots, ou les données textuelles joue un rôle important dans différentes applications  telles que l'acquisition de connaissances, les systèmes de recommandation, et le traitement du langage naturel. Au cours des dernières années, de nombreuses ontologies ont été développées et utilisées pour structurer les connaissances au sein des systèmes d'information. Le calcul de similarité sémantique à partir d'ontologie s'est développé et selon le contexte est complété par d'autres méthodes de calcul de similarité. Dans cet article, nous proposons et appliquons une approche pour le calcul de la similarité sémantique basée sur l'ontologie au sein d'un système de recommandation.
	\end{resume}
	
	\begin{motscles}
		Similarité sémantique, Ontologie, Système de Recommandation, Plongement de mots
	\end{motscles}
	
	\begin{abstract}
		Measurement of the semantic relatedness or likeness between terms, words, or text data plays an important role in different applications dealing with textual data such as knowledge acquisition, recommender system, and natural language processing. Over the past few years, many ontologies have been developed and used as a form of structured representation of knowledge bases for information systems. The calculation of semantic similarity from ontology has developed and depending on the context is complemented by other similarity calculation methods.  In this paper, we propose and carry on an approach for the calculation of ontology-based semantic similarity using in the context of a recommender system.
	\end{abstract}
	
	
	% ------------------------------------------
	% CORPS DE L'ARTICLE
	% ------------------------------------------
	
	\section{Introduction}
	Avec le développement d'Internet et du World Wide Web, les sites Web ou applications e-commerce contiennent des de données textuelles structurées, semi-structurées ou non structurées qui ne cessent d'augmenter. La recherche d'informations sur ces sources de données permet d'améliorer certaines tâches telles que la recherche, le classement. Plus précisément, le calcul de la similarité sémantique montre à quel point deux concepts, deux termes ou deux entités sont proches, sur la base de la comparaison des liens taxonomiques et des propriétés sémantiques \cite{sanchez2012ontology}.

	En structurant et en organisant un ensemble de termes ou de concepts au sein d'un domaine de manière hiérarchique et en modélisant les relations entre ces ensembles de termes ou de concepts à l'aide d'un descripteur relationnel, une ontologie permet de spécifier un vocabulaire conceptuel standard pour représenter les entités du domaine \cite{rodriguez2003determining}. Diverses applications utilisant des ontologies décrivent des termes, des entités et quantifient les relations entre eux \cite{pesquita2009semantic, jiang2011constructing}. Ces dernières années, l'utilisation d'ontologies est devenue plus populaire dans les systèmes de recommandation \cite{ibrahim2018ontology, obeid2018ontology}. Ainsi, le calcul de similarité sémantique basé sur l'ontologie permet d'améliorer la précision des tâches d'appariement, de recherche et de classement sur des éléments ou des profils d'utilisateurs.
	% recommender system
	
Une ontologie peut être représentée selon différents modèles : (1) Le modèle de représentation en triplet définit une ontologie comme un ensemble de triplets $\langle sujet, pr\acute{e}dicat, objet \rangle$   où la relation entre le sujet et l'objet est exprimée par le prédicat. Le sujet est une ressource\footnote{Une ressource peut être une classe, un instance, un concept, un nombre, un chaîne de caractères \cite{cyganiak2014rdf}}, le prédicat est une propriété d'une ressource, et l'objet identifie la valeur de la propriété de la ressource. L'objet d'un triplet peut contenir une autre ressource ou un littéral. (2) Le modèle de représentation graphique considère qu'une ontologie est un graphe orienté où les nœuds représentent les ressources ou les littéraux tandis que les arcs représentent les propriétés nommées. (3) Le modèle de représentation orienté objet définit une ontologie comme un ensemble d'objets, dans lequel les objets correspondent aux ressources et les variables d'instance de l'objet correspondent aux propriétés de ressources  \cite{cyganiak2014rdf}.
	
En considérant une ontologie comme un ensemble de triplets, les approches courantes de calcul de similarité sémantique basées sur une ontologie présentent deux points faibles. Le premier point faible concerne la mesure de similarité qui se calcule soit entre objets, soit entre objets et prédicats \cite{sanchez2012ontology, li2021ontology}. Le calcul basé sur les objets n'utilise pas les informations du sujet, alors qu'elles peuvent contenir des informations contextuelles du triplet intéressantes pour la comparaison. 
		Le second point faible concerne la distinction du type des objets : textuels ou numériques \cite{meym2016semantic}. Le calcul de similarité entre des objets numériques consiste en un simple calcul arithmétique. Le calcul de similarité entre des objets textuels est basé sur la fréquence des mots composant les objets textuels à comparer. Ce calcul ne tient pas compte de la dépendance sémantique entre ces mots. Cette dernière peut être une richesse pour la comparaison. Dans le cadre de nos travaux, nous visons le traitement de ces deux points faibles afin de définir un calcul de similarité sémantique plus précis au sein d'un système de recommandation.

	Le reste de cet article est organisé comme suit. Tout d'abord, la section 2 présente des travaux de la littérature sur lesquels s'appuie notre approche. La section 3 présente nos contributions principales sur la construction du système de recommandation exploitant la mesure de similarité entre des ensembles de triplets. Avant de conclure, nous testons nos travaux dans la section 4 à partir d'un cas expérimental traitant de l' achat/vente de véhicules d'occasion. Enfin, nous concluons et présentons les perspectives.

	\section{Travaux de la littérature}
	\subsection{Apport des ontologies}
	Dans le contexte du partage des connaissances, une ontologie est une description formelle et explicite des connaissances partagées qui consiste en un ensemble de concepts dans un domaine et les relations entre ces concepts \cite{guarino1995towards}. L'utilisation des ontologies facilite le partage et la réutilisation des connaissances entre les personnes et les applications largement diffusées. L'usage des ontologies permet \cite{abiteboul_2011} :
	\begin{itemize}
		\item L'organisation des données : une ontologie est construite sur la base des structures naturelles de l'information en permettant de visualiser les concepts et leurs relations.
		\item L'amélioration de la recherche : au lieu de rechercher par mot-clé, la recherche sur les ontologies peut renvoyer des synonymes à partir des termes de la requête. %l'utilisation d'ontologies permet d'améliorer la précision de la fonction de recherche de l'application. 
		\item L'intégration de données issues de différentes sources, différents langages.% basés sur de la colle sémantique entre des sources d'information hétérogènes afin d'offrir un point d'entrée unique à l'information.
	\end{itemize}
	%. Par conséquent, un terme dans une ontologie peut représenter un concept, un sujet, un prédicat, un
	%objet ou un ensemble de triplets.
	
	Fondamentalement, une ontologie peut être représentée par le langage OWL qui permet de contraindre les faits RDF dans un domaine particulier. Un fait RDF est défini par un triplet qui est un ensemble de trois composants : un sujet, un prédicat et un objet. Intuitivement, un triplet $\langle sujet, pr\acute{e}dicat, objet \rangle$ exprime qu'un sujet donné a une valeur donnée pour une propriété donnée\cite{abiteboul_2011, luyen2016development}. Une ontologie représentée en OWL possède un mécanisme d'inférence ou de raisonnement permettant de déduire les connaissances supplémentaires. %Ainsi, nous définons \underline{un terme} dans une ontologie peut être utilisé pour exprimer un concept, un sujet, un prédicat, un objet ou un ensemble de triplets.}
	
	La similarité sémantique basée sur l'ontologie fait référence à la proximité de deux termes\footnote{Une terme est utilisé pour exprimer un concept, un sujet, un prédicat, un objet, ou un ensemble de triplets} au sein d'une ontologie donnée. La distance entre deux termes est une représentation vectorielle numérique de la distance entre deux termes l'un de l'autre \cite{lee2008comparison}. Cela permet d'utiliser l'ontologie pour rechercher efficacement des éléments liés ou pour identifier des associations entre des termes.
	
	L'utilisation des ontologies comme une base de connaissance devient de plus en plus populaire dans les tâches de modélisation, d'inférence des nouvelles connaissances, ou de calcul de similarité pour des systèmes de recommandation \cite{du2019apports}.  Dans la section suivante, nous rappelons les notions de base des systèmes de recommandation et précisons le rôle que peut y jouer une ontologie notamment dans certains domaines.
	
	\subsection{Système de recommandation basé sur les ontologies}
	Le système de recommandation (SdR) est conventionnellement défini comme une application qui tente de recommander les éléments les plus pertinents aux utilisateurs en raisonnant ou en prédisant les préférences de l'utilisateur dans un élément en fonction d'informations connexes sur les utilisateurs, les éléments, et les intéractions entre les éléments et les utilisateurs \cite{LU201512, le2021towards}. En général, les techniques de recommandation peuvent être classées selon 6 principales approches : les SdRs basés sur les données démographiques, les SdRs basés sur le contenu, les SdRs basés sur le filtrage collaboratif, les SdRs basés sur la connaissance, les SdRs sensibles au contexte, et les SdRs hybrides.
	
	%\textcolor{red}{La quantité de données utilisées dans les expériences d'évaluation passées joue un rôle extrêmement important afin de calculer les prédictions pour les utilisateurs dans les SdRs basés sur le contenu et les SdRs basés sur le filtrage collaboratif. En fait, si la quantité de données collectées est limitée, les résultats de ces systèmes sont médiocres ou ne couvrent pas complètement le spectre des combinaisons entre les utilisateurs et les éléments. Il s'agit du problème du démarrage à froid \cite{aggarwal2016knowledge}.}
	 Dans plusieurs domaines tels que les services financiers, les produits de luxe coûteux, l'immobilier ou les automobiles, les articles sont rarement achetés et les évaluations des utilisateurs ne sont souvent pas disponibles. De plus, la description des articles peut être complexe et il est difficile d'obtenir un ensemble raisonnable de notes reflétant l'historique des utilisateurs sur un article similaire. Par conséquent, les SdRs basés sur les données démographiques, sur le contenu, et sur le filtrage collaboratif ne sont généralement pas bien adaptés aux domaines dans lesquels les éléments possèdent les caractéristiques mentionnées. 
	Des systèmes de recommandation basés sur les connaissances représentées au moyen d'ontologies sont alors proposés pour relever ces défis en sollicitant explicitement les besoins des utilisateurs pour ces éléments et une connaissance approfondie du domaine sous-jacent pour les mesures de similarité et le calcul des prédictions \cite{jannach2010}. 
	
	Pour améliorer la qualité de la recommandation, les calculs de similarité entre éléments ou le profil utilisateur dans un système de recommandation joue un rôle très important. Ils permettent d'établir une liste de recommandations tenant compte des préférences des utilisateurs obtenues suite aux déclarations des utilisateurs ou bien de leurs interactions. Nous détaillons dans la section suivante les mesures de similarité sémantique entre les éléments au sein d'un système de recommandation.
	
	\subsection{Mesure de similarité sémantique}
	Les avantages de l'utilisation des ontologies consistent en la réutilisation de la base de connaissances dans divers domaines, la traçabilité et la capacité d'utiliser le calcul et l'application à une échelle complexe et à grande échelle \cite{nguyen2011ontologies}. En fonction de la structure du contexte applicatif et de son modèle de représentation des connaissances, différentes mesures de similarité ont été proposées. En général, ces approches peuvent être classées selon quatre stratégies principales \cite{sanchez2012ontology, meym2016semantic} : (1) basée sur le chemin, (2) basée sur les caractéristiques, (3) basée sur le contenu de l'information, et (4) la stratégie  hybride qui inclut des combinaisons des trois stratégies de base.
	
	En mesurant la similarité sémantique basée sur le chemin, les ontologies peuvent être considérées comme un graphe orienté avec des nœuds et des liens, dans lequel les classes ou les instances sont interconnectées principalement au moyen de relations d’hyperonyme et d’homonyme où l’information est structurée de manière hiérarchique en utilisant la relation ‘est-un’ \cite{meym2016semantic}. Ainsi, les similarités sémantiques sont calculées en fonction de la distance entre deux classes ou instances. De cette manière, plus le chemin est long, plus les deux classes ou instances sont sémantiquement différentes \cite{sanchez2012ontology}. Le principal avantage de cette stratégie est la simplicité car elle nécessite un faible coût de calcul basé sur le modèle de graphe et ne nécessite pas les informations détaillées de chaque classe et instance \cite{li2021ontology}. Néanmoins, le principal inconvénient de cette stratégie concerne le degré de complétude, d’homogénéité, de couverture et de granularité des relations définies dans l’ontologie \cite{sanchez2012ontology}.

	Lors de la mesure des similarités sémantiques basées sur les caractéristiques, les classes et les instances dans les ontologies sont représentées comme un ensemble de caractéristiques  ontologiques \cite{sanchez2012ontology, meym2016semantic}. Les points communs entre les classes et les instances sont calculés en fonction de leur ensemble de caractéristiques ontologiques. De cette manière, l’augmentation de la différence de deux classes ou instances dépend de l’augmentation de nombreuses propriétés partagées et de la diminution des propriétés non-partagées entre elles \cite{varelas2005semantic}. 
	%\st{La définition de l’ensemble des caractéristiques est importante dans cette stratégie dans laquelle les entités peuvent être représentées comme des ensembles de caractéristiques et les informations disponibles dans les ontologies telles que l’ensemble des synonymes, les définitions et les différents types de relations sémantiques seront prises en compte. }
	L’évaluation de la similarité peut être réalisée en utilisant plusieurs coefficients sur les ensembles de propriétés tels que l’indice de Jaccard \cite{jaccard1901etude}, le coefficient de Dice \cite{dice1945}  ou l’indice de Tversky \cite{tversky1977features}. L’avantage de cette stratégie est qu’elle évalue à la fois les points communs et les différences d’ensembles de propriétés comparées qui permettent d’exploiter plus de connaissances sémantiques que l’approche basée sur le chemin. Cependant, la limitation est qu’il est nécessaire d’équilibrer la contribution de chaque propriété en décidant la standardisation et la pondération des paramètres sur chaque propriété.
	
	En mesurant les similitudes sémantiques basées sur le contenu de l'information (CI), on utilise le contenu de l'information comme mesure de l'information en associant des probabilités d'apparition à chaque classe ou instance dans l'ontologie et en calculant le nombre d'occurrences de ces classes ou instances dans l'ontologie \cite{sanchez2012ontology}.
	%\textcolor{red}{ Par exemple, le CI d'une classe ou d'une instance $c$ peut être calculé par le logarithme négatif de sa probabilité d'occurrence $p(c)$ comme suit :
	%\begin{equation}\label{equ1}
	%CI(c) = -log \; p(c)
	%\end{equation}
	%Compte tenu de cette définition de CI, Lin \cite{lin1998information} propose sa mesure de similarité sémantique comme suit :
	%\begin{equation}
	%Sim_{Lin}(c_1,c_2) = \frac{2\;.\; CI(c_{ACPI})}{CI(c_1) + CI(c_2)}
	%\end{equation}
	%où $c_1$ et $c_2$ présentent deux classes différentes dans l'ontologie, $c_{ACPI}$ est l'Ancêtre Commun le Plus Informatif de $c_1$ et de $c_2$ dans l'ontologie. 
%}
	De cette manière, les classes ou instances peu fréquentes deviennent plus informatives que les classes ou instances fréquentes. 
	Un inconvénient de cette stratégie est qu'elle exige des ontologies larges avec une structure taxonomique détaillée afin de bien différencier les classes.
	%\textcolor{red}{Cependant, cette approche nécessite des ontologies larges et claires avec une désambiguïsation et une annotation appropriées des classes et des instances. }
	
	Au-delà de la mesure des similarités sémantiques mentionnée ci-dessus, il existe un certain nombre d’approches basées sur des combinaisons des trois principales stratégies. Par exemple, Hu et al. \cite{hu2006semantic} utilisent la combinaison de la stratégie basée sur les caractéristiques et la stratégie basée sur le chemin.  Ils utilisent la logique de description pour représenter les caractéristiques des entités et la mesure de similarité cosinus pour calculer une similarité. De leur côté, Batet et al. \cite{batet2011ontology} utilisent l’équation \ref{equa1}  pour calculer la similarité sémantique basée sur les caractéristiques des classes et des instances et l’approche basée sur le contenu de l’information.
	
	\begin{equation}\label{equa1}
	Sim(c_1, c_2) = -log_{2}\frac{|T(c_1) \cup T(c_2)| - |T(c_1) \cap T(c_2)|}{|T(c_1) \cup T(c_2)|}
	\end{equation} où $T(c_i) \;=\;\{c_j \;\in C\;|\;c_j$ est la superclasse de $c_i\}$, $C$  contient la hiérarchie complète des concepts ou la taxonomie de l'ontologie.
	
	Dans nos travaux, nous avons choisi de travailler sur la représentation d'une ontologie au moyen de triplets. Un triplet RDF comporte trois composants: sujet, prédicat et objet. En particulier, le sujet peut être le nom d'une classe, ou un instance. Le prédicat est le nom d'une propriété d'un classe ou d'un instance. L'objet est une valeur d'une propriété de la classe ou du instance qui peut se séparer en un littéral ou un nom d'une autre classe ou un autre instance. Le nom d'une classe, d'un instance, ou des littéraux sont exprimés via un texte pouvant comporter plusieurs mots. Afin de préparer leur traitement, ces contenus textuels sont vectorisés. Nous précisons dans la section suivante les méthodes que nous avons étudiées à cette fin.
	
	\subsection{Représentations vectorielles de mots}\label{vecteurdemot}
Une ontologie est composée de concepts et de relations. Ces éléments sont étiquetés par des textes (un ou plusieurs mots). Pour que les machines comprennent et effectuent des calculs sur ces contenus textuels, il faut les transformer en une représentation numérique en utilisant un corpus textuel \cite{bengio2000neural}. La vectorisation de mots permet de représenter un mot par un vecteur à  valeurs réelles et ce vecteur décrit le mieux possible le sens de ce mot dans son contexte. En général, plusieurs techniques sont proposées pour vectoriser un mot telles que celles basées sur la fréquence de mots (e,g. TF-IDF \cite{salton1986introduction}) ou le sac de mots continus (CBOW) ou encore le saut de gramme (Skip-Gram) (e,g. Word2vec \cite{mikolov2013efficient}).
	
	Le TF-IDF\footnote{TF-IDF (Term Frequency-Inverse Document Frequency) est noté pour la Fréquence du Terme et la Fréquence Inverse du Document} est une mesure statistique basée sur un corpus de documents\footnote{Dans le contexte d'une ontologie, un ensemble de triplets est équivalent un document}. Cette technique évalue la pertinence d'un mot par rapport à un document dans un corpus de documents. Tout d'abord, on calcule la fréquence relative d'un mot $m$ dans un document $d$ comme suit :
	\begin{equation}\label{tf}
	tf(m,d) = \frac{f(m,d)}{\sum_{m'\in d}f(m',d)}
	\end{equation} où $f(m,d)$ dénote le nombre de fois où le mot $m$ apparaît dans le document $d$, $\sum_{m'\in d}f(m',d)$ dénote le nombre total des mots dans le document $d$. Ensuite, on mesure la quantité d'informations fournies par le mot $m$ dans le corpus de documents $D$ avec la fréquence inverse du document comme suit :
	\begin{equation}\label{idf}
	idf(m,D) = log\frac{N}{|d \in D:m\in d|} + 1
	\end{equation}
	où $N$ est le nombre de documents dans le corpus, $|d \in D:m\in d|$ est le nombre de documents où le mot $m$ apparaît. Donc, la valeur de $tf.idf$ du mot $m$ dans le document $d$ au sein du corpus $D$ est définie comme suit :
	\begin{equation}\label{tfidf}
	tf.idf(m,d,D) = tf(m,d)\times idf(m,D)
	\end{equation}
	Une valeur $tf.idf(m,d,D)$ élevée d'un mot $m$ dans un document $d$ indique que ce mot est pertinent pour ce document au sein du corpus de documents $D$ \cite{salton1986introduction}.
	
	La technique de sac de mots continus, CBOW, construit la représentation vectorielle d'un mot $m_i$ via la prédiction de son occurrence et la connaissance des mot avoisinants. Autrement dit, le saut de gramme, Skip-Gram, construit la représentation vectorielle d'un mot $m_i$ en prédisant son contexte d'occurrence. Donc, étant donné une séquence de mots d'apprentissage $\{ m_1, m_2, ..., m_T\}$ l'objectif du CBOW est de maximiser la moyenne des log-probabilités :
	\begin{equation} 
	\frac{1}{T}\sum_{t=1}^{T}\sum_{-c\leq j\leq c, j \neq 0 }log\; p(m_{t}|m_{t+j})
	\end{equation}
	Tandis que l'objectif de Skip-gram est de maximiser la moyenne des log-probabilités :
	\begin{equation} 
	\frac{1}{T}\sum_{t=1}^{T}\sum_{-c\leq j\leq c, j \neq 0 }log\; p(m_{t+j}|m_{t})
	\end{equation} où $c$ est la taille du contexte. La formulation de Skip-gram définit $p(m_{t+j}|m_t)$ en utilisant la fonction softmax :
	\begin{equation}
	p(m_{t+j}|m_t) = \frac{exp((v'_{m_{t+j}})^T v_{m_{t}})}{\sum_{i=1}^{M} exp((v'_{m_{i}})^T v_{m_{t}})}
	\end{equation} où $v_{m_t}$ est la représentation vectorielle d'entrée du mot $m_t$, et $v'_{m_{t+j}}$, $v'_{m_i}$ sont les représentation vectorielles de sortie du mot $m_{t+j}$, $m_{i}$. $M$ est le nombre de mots dans le dictionnaire du corpus. 
	
\begin{figure}[h!]
		\begin{center}
			%left, bottom, right, top
			\includegraphics[width=0.48\textwidth]{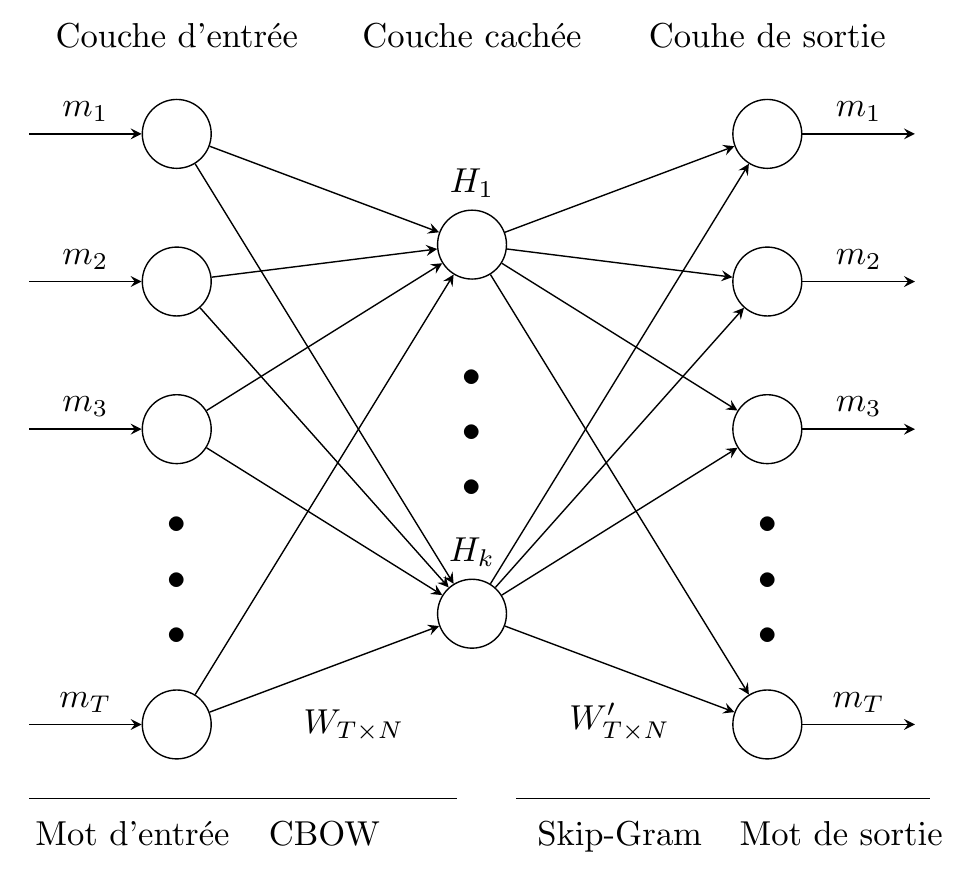}
		\end{center}
		\caption{L'architecture du modèle de Word2vec}
		\label{fig00}
	\end{figure}

	Word2vec est l'une des implémentations les plus populaires pour créer un plongement de mots en utilisant une architecture d'apprentissage automatique à l'aide d'un réseau de neurones. Il prédit les mots en fonction de leur contexte en combinant les deux techniques CBOW et Skip-gram \cite{mikolov2013efficient, ahmia2019utilite}. En particulier, la figure \ref{fig00} illustre l'architecture  de Word2vec qui comporte conventionnement trois couches : couche d'entrée, couche cachée, et couche de sortie. D'abord, un dictionnaire de mots avec la taille $N$ est synthétisé à partir d'un corpus de textes. Ensuite, le processus d'apprentissage automatique crée et met à jour les valeurs des poids des matrices $W_{T\times N}$, $W'_{T\times N}$. Une fois l'apprentissage terminée, nous obtenons la matrice $W_{T\times N}$ pour le plongement de mots.
	
		%Cette architecture comporte conventionnement trois couches: couche d'entrée, couche cachée, et couche de sortie. La couche d'entrée  Lqui permet d'obtenir les représentation vectorielle à partir d'un grand corpus d'entrée \cite{mikolov2013efficient, ahmia2019utilite}

	Plusieurs plongements de mots sont créés en utilisant ce modèle pour des langues différentes \cite{mikolov2013efficient}. Fauconnier \cite{fauconnier_2015}, et Hadi et ses collègues \cite{abdine2021evaluation} implémentent ce modèle à partir des textes en français. D'ailleurs, plusieurs autres travaux ont obtenu de bons résultats dans la conversion d'un mot en  une représentation vectorielle tels que Fastext \cite{fastexr}, Glove \cite{pennington2014glove},  le modèle Transformer avec l'implémentation du BERT \cite{devlin2018bert}.
	
	Un plongement de mots entraîné avec de très grands corpus permet d'obtenir rapidement la représentation vectorielle d'un mot. Dans nos travaux, nous avons fait le choix de calculer la mesure de similarité entre deux termes textuels en tenant compte de la combinaison de CBOW et Skip-gram. La similarité entre les deux termes textuels qui se composent de mots différents peut profiter de cette forme de représentation afin de calculer la distance entre eux. Dans la section suivante, nous détaillons notre approche proposée pour mesurer de similarité au sein d'un système de recommandation.
	
	\section{Mesure de similarité au sein d'un système de recommandation}
	\subsection{Système de recommandation pour l'achat/vente des véhicules d'occasion}
	Dans le cadre de nos travaux, nous nous intéressons à l'illustration de la mesure de similarité sémantique sur le système de recommandation basé sur les connaissances représentées au moyen d'ontologies dans une application e-commerce de vente/achat des véhicules d'occasion.
	%\textcolor{red}{Dans le cadre de notre travaux, nous nous intéressons sur le système de recommandation basée sur les connaissances représentées au moyen d'ontologies dans une application e-commerce de vente/achat des véhicules d'occasion.} illustrer
	
	Les données d'un SdR basé sur la base de connaissances représentées au moyen d'ontologies se concentrent sur trois types principaux : les profils de l'utilisateur, les descriptions d'éléments ou les attributs d'éléments, et les intéractions entre les utilisateurs et les éléments. Tout d'abord, les profils d'utilisateur incluent les informations personnelles et les préférences de l'utilisateur sur les éléments de véhicule. Ils peuvent être organisés et être réécrits sous la forme des triplets formellement définis comme suit :
	\begin{equation}
	G_U = \{ a_1^u, a_2^u, ...,  a_n^u,  \}
	\end{equation}
	où $a_i^u$ dénote le triplet $a_i^u = \langle sujet_i, pr\acute{e}dicat_i, objet_i \rangle$. Autrement dit, le triplet $a_i^u$ peut aussi s'exprimer comme $\langle ressource_i, propri\acute{e}t\acute{e}_i, \acute{e}nonc\acute{e}_i \rangle$. Par exemple, ``\textit{Louis aime la voiture modèle S de Tesla}''. Cette expression naturelle peut se représenter sous la forme de deux triplets différents comme $\langle Louis, aime, la\_voiture\_mod\grave{e}le\_s \rangle$, $\langle la\_voiture\_mod\grave{e}le\_s, est\_fabiqu\acute{e}e\_par, Tesla \rangle$. Ensuite, les descriptions de véhicule peuvent également être représentées comme un graphe de connaissance. Elles peuvent être définies selon la même approche : 
	\begin{equation}
	G_V = \{ a_1^v, a_2^v, ...,  a_n^v,  \}
	\end{equation} où $a_i^v$ dénote le triplet $a_i^v = \langle sujet_i, pr\acute{e}dicat_i, objet_i \rangle$ ou $a_i^v = \langle ressource_i, propri\acute{e}t\acute{e}_i, \acute{e}nonc\acute{e}_i\rangle$. Enfin, lorsqu'un utilisateur effectue une intéraction sur des éléments de description de véhicule en donnant une note, un commentaire ou en ajoutant à une liste de favoris, on marque ces intéractions pour avoir une analyse de l'intention et du comportement de l'utilisateur afin de proposer des recommandations pertinentes. Donc, les intéractions sont définies comme une fonction à plusieurs paramètres :
	\begin{equation}
	SR : G_U \times G_V \times G_{C_1} \times ... \times G_{C_n} \rightarrow Int\acute{e}raction
	\end{equation} où $G_U$ correspond à l'utilisateur, $G_V$ correspond aux éléments de description de véhicule, $G_{C_i}$s correspond aux informations contextuelles, par exemple : objectifs, locations, temps, ressources \cite{Adomavicius2011}.   
	Les ontologies sont développées pour profiler des utilisateurs et modéliser des éléments de description de véhicules \cite{le2021towards}. Sur la base de ces ontologies, les données RDFs sont collectées et stockées dans un triplestore interrogeable au moyen de requêtes SPARQL. Des règles peuvent être définies pour déduire ou filtrer les éléments en utilisant les inférences ontologies. Dans ce cas, le SdR basé sur les connaissances comporte les quatre principales tâches suivantes : 
	\begin{itemize}
		\item Recevoir et analyser les demandes des utilisateurs à partir de l'interface utilisateur.
		\item Construire et réaliser des requêtes sur la base de connaissance.
		\item Calculer des similarités sémantiques entre l'élément de véhicules, le profil utilisateur.
		\item Classer les éléments correspondant aux besoins de l'utilisateur.
	\end{itemize}
	
	% How semantic similiarty is used in the RS, example
	Les mesures de similarité entre les éléments ou le profil utilisateur est une tâche importante pour générer la liste des recommandations la plus pertinente. Le travail s'effectue à partir des données RDFs qui sont organisées sous la forme de triplets $\langle sujet, pr\acute{e}dicat, objet \rangle$. Les comparaisons entre deux triplets se limitent souvent aux objets communs ou non communs. Les informations de sujet et prédicat peuvent cependant également fournir des informations importantes sur l'objet lui-même et sa comparaison avec d'autres triplets. Dans la section suivante nous présentons comment dans notre approche nous exploitons ces deux accès à l'information pour calculer les similarités sémantiques entre les triplets d'une base de connaissances.

	\subsection{Mesure de similarité sémantique entre les triplets}
	%La revue de la littérature ci-dessus donne un aperçu de la similarité sémantique basée sur l'ontologie existante. Chaque méthode a ses avantages et inconvénients.
	 Nous avons choisi de définir une approche hybride tenant compte de la combinaison des approches de calcul de la mesure de similarité sémantique basées sur les caractéristiques et basées sur le contenu de l'information. Le sujet, le prédicat et l'objet dans un triplet contiennent des informations importantes. Un ensemble de triplets permet d'agréger des informations provenant de triplets simples. Par conséquent, la mesure de la similarité sémantique entre ensembles de triplets doit prendre en compte tous les triplets/éléments de chaque ensemble.
	
	La mesure de la similarité sémantique se concentre sur la comparaison de deux ensembles de triplets à partir de tous leurs éléments en les séparant en informations quantitatives et informations qualitatives. D'une part, la comparaison d'objets est réalisée en utilisant la stratégie de similarité sémantique basée sur les propriétés. D'autre part, la comparaison des sujets et des prédicats est effectuée par la stratégie de similarité sémantique basée sur le contenu de l'information.
	\subsection{Mesure des informations qualitatives }
	Les informations qualitatives font référence aux mots, aux étiquettes utilisés pour décrire les classes, les relations, et les annotations. Dans un triplet, le sujet et le prédicat expriment une information qualitative. Les objets peuvent contenir des informations qualitatives ou quantitatives. Par exemple, nous avons trois triplets suivants : 
	$\langle ford\_focus\_4\_2018, la\_bo\hat{\I}te\_de\_vitesse, m\acute{e}canique \rangle$
	
	$\langle ford\_focus\_4\_2020, la\_bo\hat{\I}te\_de\_vitesse, m\acute{e}canique \rangle$
	
	$\langle citro\"{e}n\_c5\_aircross, la\_bo\hat{\I}te\_de\_vitesse, m\acute{e}canique \rangle$ 
	
	Tous les composants de ces trois triplets sont qualitatifs. L'information du sujet de trois triplets peut être utilisée pour contribuer à la mesure de similarité entre eux. Dans cette section, nous nous concentrons sur la mesure de la similarité sémantique pour les Sujets, Prédicats et Objets Qualitatifs (SPOQ). Nous proposons la même formule pour les trois composants afin de calculer la similarité.
	
	Soient deux SPOQs $a_{s1}$ et $a_{s2}$ dont les vecteurs de mots sont $M_1 = \{m_{11}, m_{12}, ..., m_{1k}\}$ et $M_2 = \{m_{21}, m_{22}, ..., m_{2l}\}$, leur similarité sémantique est définie comme suit :
	\begin{equation}
	Sim_1(a_{s1},a_{s2}) = \frac{\sum_{i=1}^{k} \bar{S}(m_{1i}, a_{s2}) + \sum_{j=1}^{l} \bar{S}(m_{2j}, a_{s1})}{k + l}
	\end{equation}
	où $\bar{S}(m, a_{s})$ dénote la similarité sémantique d'un mot $m$ et d'un SPOQ. La fonction $\bar{S}(m, a_{s})$ est formellement calculée  comme suit :
	\begin{equation}
	\bar{S}(m, a_{s}) = \max\limits_{m_i \in M} \bar{S}(m, m_i)
	\end{equation} où $m_i \in M=\{m_1, m_2, ..., m_k\}$ est le vecteur de mots de $a_s$. Chaque mot $m_i$ est représenté par un vecteur numérique. On peut utiliser les techniques introduits dans le section \ref{vecteurdemot}. L'approche basée sur la fréquence de mots TF-IDF facilite l'obtention de la probabilité d'un mot dans un ensemble de triplets. Cependant, le principal inconvénient de cette approche est qu'elle ne peut pas capturer l'information sémantique du mot et l'ordre du mot dans l'ensemble de triplets parce qu'elle crée le vecteur basé sur la fréquence du mot dans un ensemble de triplets et la collection des ensembles de triplets. Nous proposons l'utilisation de CBOW et Skip-gram avec l'implémentation de Word2vec \cite{mikolov2013efficient, abdine2021evaluation} afin de surmonter cela. Nous calculons finalement la similarité entre deux mots $m_i$, $m_j$ par la similarité cosinus : $\bar{S}(m_i, m_j) = \frac{m_i . m_j}{\|m_i\|\|m_j\|}$.
	\subsection{Mesure des informations quantitatives}
	Les informations quantitatives sont des informations numériques qui sont utilisées pour exprimer l'information de type nominal, ordinal, intervalle, ou ratio. Dans un triplet, l'objet utilise souvent cette forme d'information pour manifester des informations des propriétés pour les classes, concepts de l'ontologie. Par exemple, nous avons des triplets suivants :
	$\langle ford\_focus\_4\_2018, a\_le\_kilom\acute{e}trage, 107351\rangle$
	
	$\langle ford\_focus\_4\_2020, a\_le\_kilom\acute{e}trage, 25040 \rangle$
	
	$\langle citro\"{e}n\_c5\_aircross, a\_le\_kilom\acute{e}trage, 48369 \rangle$ 
	
	Les objets de ces triplets sont des valeurs numériques. La comparaison entre chiffres s'effectue simplement par les mesures de distance.
	Afin de comparer deux objets différents, nous utilisons la distance euclidienne entre deux objets. Ainsi, plus la différence entre deux objets est petite, plus la similitude entre eux est grande. Soient deux objets $a_{o1}$ et $a_{o1}$ dont les vecteurs sont $a_{o1} = \{o_{11}, o_{12}, ..., o_{1k}\}$ et $a_{o2} =\{o_{21}, o_{22}, ..., o_{2k}\}$, leur similarité sémantique est définie comme suit :
	
	\begin{equation}
	Sim_2(a_{o1},a_{o2}) = \frac{1}{1 + \sqrt{\sum_{i=0}^{k}(o_i - o_j)^2}}
	\end{equation}
	\subsection{Mesure des triplets}
	La comparaison de deux triplets $a_1=\langle a_{s1}, a_{p1}, a_{o1}  \rangle$ et $a_2=\langle a_{s2}, a_{p2}, a_{o2}  \rangle$ est effectuée en fonction du type d'information des objets dans les triplets. Si l'objet contient des informations qualitatives, la similarité sémantique entre $a_1$ et $a_2$ est définie comme suit :
	\begin{equation}
	Sim_{I} (a_1,a_2) = \frac{1}{N}\sum_{i \in P, \omega \in Q} \omega \times Sim_1(a_{i1},a_{i2}) 
	\end{equation} où $P = \{s, p, o\}$ correspond aux informations de $sujet$, $predicat$, et $objet$ sous la forme de vecteur de mots. $Q = \{\alpha, \beta, \gamma\}$ est le poids respectifs pour les composants de triplet. $N$ est le nombre de composants de triplet.
	
	Par ailleurs, si l'objet contient des informations quantitatives, la mesure de similarité sémantique des triplets $a_1$ et $a_2$ est définie comme suit :
	
	\begin{equation}
	\begin{split}
	Sim_{II} (a_1,a_2) =  \frac{1}{N} (\sum_{i \in P, \omega \in Q} \omega \times Sim_1(a_{i1},a_{i2}) + \\ \gamma \times Sim_2(a_{o1},a_{o2}))
	\end{split}
	\end{equation} où $P = \{s, p\}$ correspond aux informations de $sujet$ et $predicat$ sous la forme de vecteur de mots. $Q = \{\alpha, \beta\}$ représente les poids respectifs du sujet et du prédicat. Et $\gamma$ est le poids pour l'objet. 
	
	Par conséquent, la similarité sémantique de deux ensembles de triplets $G_1=\{a_1, a_2, ..., a_g\}$ et $G_2=\{a_1, a_2, ..., a_g\}$ est calculée sur la base de comparaison de similarité de chaque triplet simple comme suit :
	\begin{equation}
	\begin{split}
	Sim(G_1, G_2) = \frac{1}{L}(\sum_{j=0}^{L} Sim_{I}(a_{1j},a_{2j})) \;+ \\ \frac{1}{H}(\sum_{j=0}^{H}Sim_{II}(a_{1j},a_{2j})) 
	\end{split}
	\end{equation} où $L$ est le nombre de triplets qui contient les objets qualitatifs. $H$ est le nombre de triplets qui contient les objets quantitatifs.
	
	\section{Cas expérimental}
\begin{figure}[h!]
		\begin{center}
			%left, bottom, right, top
			\includegraphics[width=0.52\textwidth]{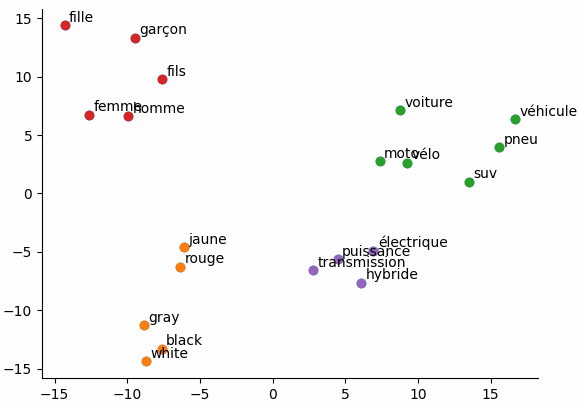}
		\end{center}
		\caption{La distance proche entre les mots, groupes qui sont vectorisés par le corpus de mots entraîné}
		\label{fig01}
	\end{figure}

	Dans cette section nous testons notre approche dans le cas d'une application d'achat/vente de véhicules. Nous mesurons ainsi la similarité sémantique entre deux ensembles de triplets représentant chacun un véhicule. Tout d'abord, la transformation d'un mot à un vecteur est réalisée en utilisant le corpus de mots entraîné qui est développé dans le travail de Hadi et al \cite{abdine2021evaluation}. Nous avons choisi d'utiliser le modèle CBOW et Skip-gram au lieu de TF-IDF à cause des problèmes concernant la capture de l'information sémantique qui est presque impossible sur la technique de TF-IDF. La figure \ref{fig01} démontre la distance très proche des mots, groupes de mots dans un même secteur en utilisant les vecteurs entraînés de Word2vec.
	
\begin{figure}[h!]
		\begin{center}
			%left, bottom, right, top
			\includegraphics[width=0.48\textwidth]{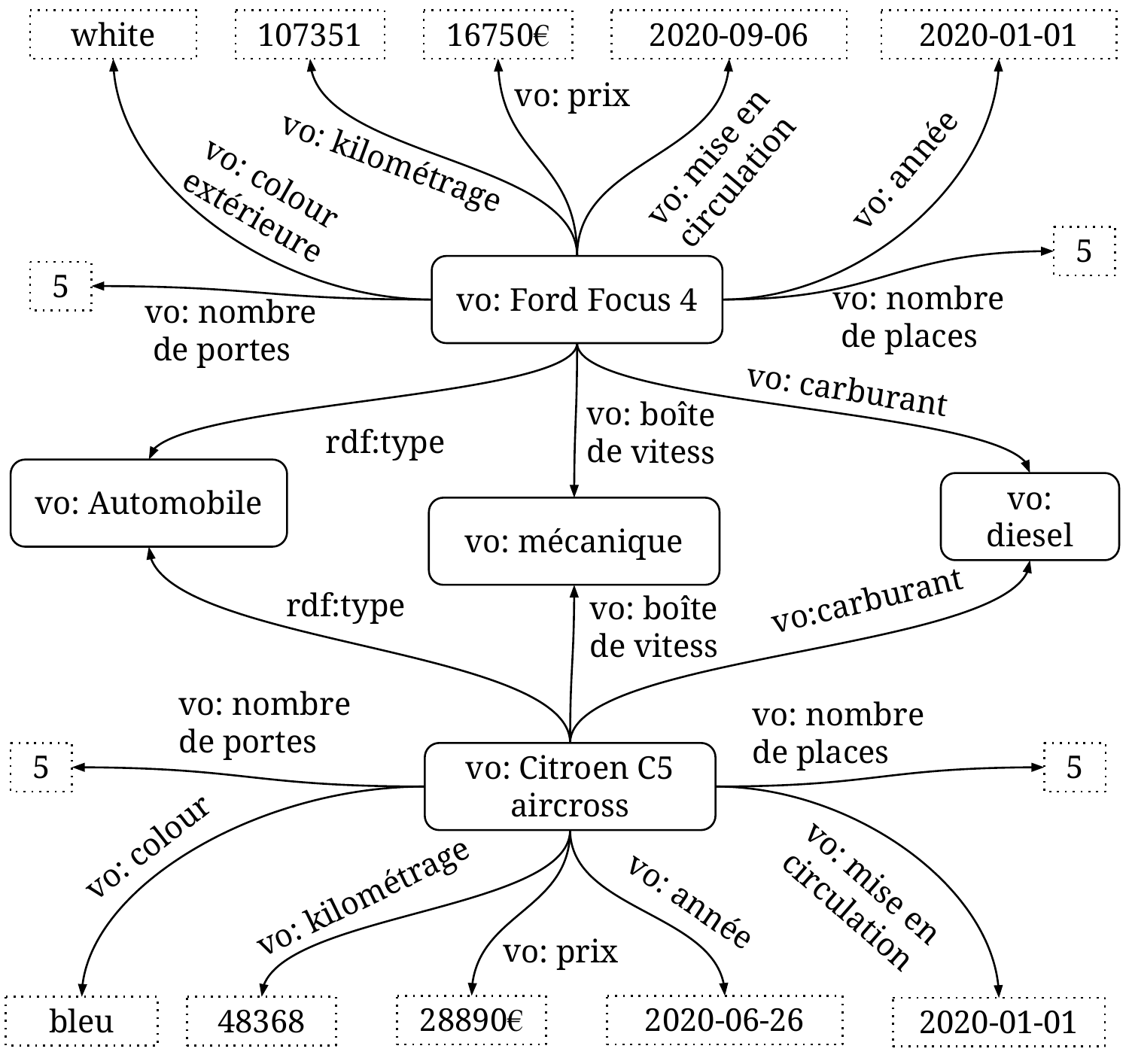}
		\end{center}
		\caption{Les données de triplets sont représentées par un graphe (\textit{vo} note pour l'Ontologie de Véhicule)}
		\label{fig02}
	\end{figure}

	En utilisant l'ontologie, nous pouvons reconstruire la base de connaissances d'un domaine sous une forme lisible par des machines ainsi que les humaines. À partir des ontologies des véhicules développées dans le travail \cite{le2021towards}, nous réalisons une collection des instances des classes et leurs relations afin de créer un triplestore de données RDF. La figure \ref{fig02} illustre deux ensembles de triplets représentant deux voitures. 
	%La figure \ref{fig02} montre aussi que les données de triplets peuvent être visualisées par un graphe où les sujets ou objets sont des nœuds et les prédicats sont les arcs.  
	
	La description de chaque véhicule est représentée par les informations textuelles et les informations numériques. Notre approche propose de séparer la mesure de la similarité en deux calculs séparés. L'un s'applique aux informations numériques parce qu'elles exigent les calculs simples pour avoir la distance ou la similarité.  L'autre s'applique aux informations textuelles lorsqu'une classe où un instance d'une ontologie se composent à partir d'un groupe de mots et chaque mot a des dépendance sémantique avec les autres. En profitant des travaux du domaine de traitement du langage naturel avec les méthodes d'apprentissage profond sur de très grands corpus, les données catégorielles peuvent être représentées dans un vecteur numérique qui contient les relations du mot avec les mots récurant provenant de plusieurs documents en ligne.
	
	\begin{table}[h!]
		\begin{tabular}{| P{0.2cm}| P{0.6cm}| P{0.9cm} | P{0.9cm} | P{0.9cm} | P{0.9cm} | P{0.9cm} |}
			\hline	
			%\multicolumn{1}{|c}{\textbf{Voiture I}} & \multicolumn{1}{|c|}{Voiture II} \\\hline
			\multicolumn{2}{|c|}{}& $V_1$ & $V_2$ & $V_3$ & $V_4$ & $V_5$ \\\hline
			\multirow{3}{*}{$V_1$} & SiLi & 1.0 &  &  & &  \\\cline{2-3}
			& N-2 & 1.0 &  &  & &  \\\cline{2-3}
			& \textbf{N-1} & 1.0 &  &  & & \\\cline{1-4}
			\multirow{3}{*}{$V_2$} & SiLi & 0.57 & 1.0 &  & &  \\\cline{2-4}
			& N-2 & 0.50  & 1.0 &  & &  \\\cline{2-4}
			& \textbf{N-1} & \textbf{0.61} & 1.0  &  & & \\\cline{1-5}
			\multirow{3}{*}{$V_3$} & SiLi & 0.50 & 0.48 & 1.0  & &  \\\cline{2-5}
			& N-2 & 0.48  & 0.46 & 1.0 & &  \\\cline{2-5} 
			& \textbf{N-1} & \textbf{0.64}  & \textbf{0.57}  & 1.0  & & \\\cline{1-6}
			\multirow{3}{*}{$V_4$} & SiLi & 0.54 & 0.49 & \textbf{0.62}  & 1.0 &  \\\cline{2-6}
			& N-2 & 0.49 & 0.47 & 0.50 & 1.0 &  \\\cline{2-6} 
			& \textbf{N-1} & \textbf{0.58}  & \textbf{0.60}   & 0.59   & 1.0 & \\\hline
			\multirow{3}{*}{$V_5$} & SiLi & 0.54  & 0.46 & \textbf{0.69 } & 0.68 & 1.0   \\\cline{2-7}
			& N-2 & 0.48  & 0.45 & 0.53 & 0.52 & 1.0   \\\cline{2-7}
			& \textbf{N-1} & \textbf{0.59}  & \textbf{0.55}   & 0.60   & \textbf{0.71}  & 1.0 \\\hline
			
			%\multirow{3}{c}{$V_1$} & \\ \cline{2-4} & Nos-1 & -2 & SiLi\\

			%$V_1$ & 1.0 &  &  & & \\\hline
			%$V_2$ & 0.61 & 1.0 &  & & \\\hline
			%$V_3$ & 0.64 & 0.57 & 1.0 & & \\\hline
			%$V_4$ & 0.58 & 0.60 & 0.59  & 1.0 & \\\hline
			%$V_5$ & 0.59 & 0.55  & 0.60  & 0.71  & 1.0 \\\hline		 
		\end{tabular}
		\centering
		\caption{La mesure de similarité entre les 5 voitures avec les trois approches différentes}
		\label{table01}
	\end{table}
	Sur la base des instances collectées, nous réalisons des expérimentations et des évaluations sur trois approches suivantes :
	\begin{enumerate}
		\item N-1 : notre approche proposée principale avec l'utilisation du modèle de Word2vec \cite{abdine2021evaluation} pour vectoriser les informations qualitatives.
		\item N-2 : notre approche avec l'utilisation du modèle de TF-IDF pour vectoriser les informations qualitatives.
		\item SiLi : l'approche proposée par Siying Li et ses collèges \cite{li2021ontology}, cette approche hybride combine la stratégie basée sur le contenu et celle sur les caractéristiques mais ne considère que les objets et les prédicats des triplets. 
			
			%\textcolor{red}{Nous choisissons ce travail pour comparer avec nos travaux parce qu'il est l'un des travaux les plus récents sur la mesure de similarité sémantique sur l'ontologie. En particulier, cette approche utilise la même stratégie avec nos travaux qui est basée la stratégie hybride  en combinant la stratégie basée sur les caractéristiques et basée sur le contenu de l'information. Différent de notre approche, SiLi propose l'utilisation des objets et des prédicats tandis que notre approche propose d'utiliser les trois composants d'un triplet : sujet, prédicat et objet.}
		
		%pour avoir des comparaisons avec les nos modèles . \textcolor{red}{dire en quoi consiste cette approche, quelles différences avec N-1 et N-2}

	\end{enumerate}
	%\textcolor{red}{ D'abord, nous effectuons  (N-1) pour obtenir les représentation vectorielles pour les mots. Ensuite, nous expérimentons  (N-2) pour obtenir les représentation vectorielles pour les mots au lieu de l'utilisation du Word2Vec. Nous implémentons  pour avoir des comparaisons avec les nôtres \cite{li2021ontology}.}
	
	La table \ref{table01} affiche les résultats de calcul de la similarité entre les 5 voitures de marques différentes. En particulier, $V_1$ est le ``\textit{Renault captur 2}'', $V_2$ est la marque ``\textit{posrche taycan}'', $V_3$ est le ``\textit{ford focus 4}'', $V_4$ est la marque ''\textit{audi a1 sportback}``, et $V_5$ est le ``\textit{citroen c5 aircross}''. Les ensembles de triplets de ces voitures sont montrés dans l'appendice \ref{appendice}. 
	
	En analysant les résultats obtenus et présentés dans la table \ref{table01}, nous arrivons sur plusieurs conclusions. Premièrement, notre approche \textbf{N-1} donne le résultat de calcul de la similarité entre les voitures plus élevé que les autres approches dans 8 sur 10 cas de comparaisons. Toutefois, le résultat de calcul de la similarité de notre approche est moins élevé que l'approche de \textbf{SiLi} dans la comparaison de deux cas : entre les voitures $V_4$, $V_3$ et entre les voitures  $V_5$, $V_3$. Deuxièmement, notre approche en utilisant le technique TF-IDF \textbf{N-2} pour la représentation vectorielle de mot a obtenu les résultats le plus bas dans tout les cas de comparaison. Cela s'explique par la capacité de capture des informations contextuelles et sémantiques de l'approche Word2vec qui est meilleur que celle de l'approche TF-IDF.
	
	% Cela est expliqué par la capacité de capture des informations contextuelles et sémantiques des approches de Word2Vec est meilleurs de l'approche TF-IDF.
	
	%l'utilisation de Word2vec permet d'obtenir de meilleurs résultats sur presque les comparaisons dans la table. Cela est expliqué par la capacité de capture des informations contextuelles et sémantiques des approches de Word2Vec est meilleurs de l'approche TF-IDF. Deuxièmement, notre approche proposée a obtenu les meilleurs résultats dans la majorité des cas en comparant avec l'approche de Siying Li. Donc, la proposition de l'utilisation de tous les composants dans un triplet est une des raisons pour ces résultats.
	
	Les expérimentations montrent que notre approche \textbf{N-1} a obtenu de bons résultats pour les mesures de similarité entre les ensembles de triplets. L'utilisation du sujet dans la comparaison permet d'ajouter de l'information à la mesure de similarité d'un triplet. Aussi, la distinction contenus textuels et numériques permet d'appliquer la formule appropriée selon le type de contenu. Au final la somme des deux calculs représente la similarité mesurée. Compte tenu des cette distinction, de la prise en compte des triplets contextuels et du calcul à partir des contenus textuels enrichi des dépendances sémantiques entre les mots constituant le texte, la similarité obtenue est plus précise que celles rencontrées dans la littérature \cite{sanchez2012ontology, meym2016semantic, li2021ontology}.
	
	%\textcolor{red}{\textbf{Aussi, la distinction des contenus textuels et numériques aide à s'appliquer la formule appropriée pour calculer chaque type de contenues. La performance des calcul des contenus numériques est améliorée parce que ces calculs sont des opérations arithmétiques ordinaires. Lors que la comparaison entre deux contenus textuels inclut plus d'informations. En particulier, les dépendances sémantiques d'un mot aux autre mots dans les contenus textuels sont capturées et représentées par un vecteur numérique. La similitude entre deux contenues textuels illustre par la petite distance entre leurs vecteurs numériques.}}
	
	% par le modèle d'apprentissage utilisant les deux technique CBOW et Skip-gram.}
	
	%\textcolor{red}{ Aussi, la séparation de calculs des informations textuelles et numériques aide améliorer la performance de calculs sur les information numériquesAlors que l tandis que les informations textuelles sont transformées en une représentation vectorielles numériques à partir du modèle d'apprentissage et un grand corpus de texte en français. }
	
	\section{Conclusion et perspectives}
	La mesure de similarité sémantique sur la base de l'ontologie est une tâche importante pour proposer une liste de recommandations pertinentes à un utilisateur. Dans cet article, nous proposons une stratégie hybride qui combine la stratégie basée sur les caractéristiques et basée sur le contenu de l'information. Avec notre approche, afin de ne pas perdre d'information, les trois composants d'un triplet sont considérés dans le calcul de similarité. La distinction de type de données, textuel ou numérique, permet d'effectuer un calcul adapté et plus précis. Nous avons effectué une première expérimentation de notre approche et l'avons comparée à deux autres calculs de similarité. Les résultats obtenus montrent son intérêt. Nous devons maintenant poursuivre nos travaux et en premier lieu effectuer d'autres tests sur des corpus différents et des applications différentes. 
	 Nous devons concéder que les mots non considérés dans le corpus entraîné posent un problème. En perspective, la résolution de ce problème ainsi que la construction d'un corpus des triplets entraînés pourraient être des travaux prometteurs dans le futur.
	%\section*{Remerciements}
	%Les remerciements s'expriment juste avant la bibliographie, à l'aide d'une Section non numérotée.

	\bibliographystyle{plain}
	\bibliography{bibliotheque} 
	
	\appendix
	\section{Appendice : Ensembles de triplets des voitures utilisés dans les expérimentations}\label{appendice}
	\lstset{ literate={á}{{\'a}}1 {é}{{\'e}}1 {í}{{{\'\i}}}1 {ó}{{\'o}}1 {ô}{{\^{o}}}1 {î}{{\^{i}}}1 {ö}{{\"o}}1 {ő}{{\H o}}1 {ú}{{\'u}}1 {Ú}{{\'U}}1 {ü}{{\"u}}1 {ű}{{\H u}}1 {Ü}{{\"U}}1 }
	
\begin{lstlisting}[captionpos=b, caption=$V_1$ Renault Captur 2 (RC2), label=lst:v1,
	basicstyle=\ttfamily\small,breaklines=true,frame=single]
 <vo:RC2,rdf:type,vo:Automobile>
 <vo:RC2,vo:année,vo:2022-01-01>
 <vo:RC2,vo:mis en circulation,vo:2022-04-28>
 <vo:RC2,vo:contrôle technique,vo:non requis>
 <vo:RC2,vo:kilométrage,vo:5493>
 <vo:RC2,vo:carburant,vo:hybride essence électrique>
 <vo:RC2,vo:boîte de vistesse,vo:automatique>
 <vo:RC2,vo:coleur extérieure,vo:noir>
 <vo:RC2,vo:nombre de portes,vo:5>
 <vo:RC2,vo:nombre de places,vo:5>
 <vo:RC2,vo:puissance fiscale,vo:5>
 <vo:RC2,vo:puissance din,vo:93>
 <vo:RC2,vo:Critique d'Air,vo:1>
 <vo:RC2,vo:emission de CO2,vo:35>
 <vo:RC2,vo:consommation mixte,vo:1.5>
 <vo:RC2,vo:norme euro,vo:euro6>
 <vo:RC2,vo:fabriquer par,vo:Renault occasion>
 <vo:RC2,vo:type de véhicule,vo:4x4, SUV & Crossover occasion>
 <vo:RC2,vo:location,vo:Cher>
 <vo:RC2,vo:price,vo:36580>
\end{lstlisting}
	
\begin{lstlisting}[captionpos=b, caption=$V_2$ Porsche Taycan (PT), label=lst:v2,
	basicstyle=\ttfamily\small,breaklines=true,frame=single]
 <vo:PT,rdf:type,vo:Automobile>
 <vo:PT,vo:année,vo:2022-01-01>
 <vo:PT,vo:mis en circulation,vo:2022-09-10>
 <vo:PT,vo:contrôle technique,vo:non requis>
 <vo:PT,vo:kilométrage,vo:4932>
 <vo:PT,vo:carburant,vo:electrique>
 <vo:PT,vo:boîte de vistesse,vo:automatique>
 <vo:PT,vo:coleur intérieure,vo:cuir ivoire>
 <vo:PT,vo:coleur extérieure,vo:noir metal>
 <vo:PT,vo:nombre de portes,vo:4>
 <vo:PT,vo:nombre de places,vo:4>
 <vo:PT,vo:garranty,vo:20 mois>
 <vo:PT,uvso:puissance fiscale,vo:8>
 <vo:PT,vo:puissance din,vo:530>
 <vo:PT,vo:Critique d'Air,vo:0>
 <vo:PT,vo:emission de CO2,vo:0>
 <vo:PT,vo:norme euro,vo:euro6>
 <vo:PT,vo:fabriquer par,vo:Porsche occasion>
 <vo:PT,vo:type de véhicule,vo:Berline occasion>
 <vo:PT,vo:location,vo:Rhône>
\end{lstlisting}

\begin{lstlisting}[captionpos=b, caption=$V_3$ Ford Focus 4 (FF4), label=lst:v3,
	basicstyle=\ttfamily\small,breaklines=true,frame=single]
 <vo:FF4,rdf:type,vo:Automobile>
 <vo:FF4,vo:année,vo:2020-01-01>
 <vo:FF4,vo:mis en circulation,vo:2020-09-06>
 <vo:FF4,vo:contrôle technique,vo:non requis>
 <vo:FF4,vo:kilométrage,vo:107351>
 <vo:FF4,vo:carburant,vo:diesel>
 <vo:FF4,vo:boîte de vistesse,vo:mécanique>
 <vo:FF4,vo:coleur extérieure,vo:gris foncé>
 <vo:FF4,vo:nombre de portes,vo:5>
 <vo:FF4,vo:nombre de places,vo:5>
 <vo:FF4,vo:garranty,vo:12 mois>
 <vo:FF4,vo:puissance fiscale,vo:4>
 <vo:FF4,vo:puissance din,vo:95>
 <vo:FF4,vo:Critique d'Air,vo:2>
 <vo:FF4,vo:emission de CO2,vo:89>
 <vo:FF4,vo:consommation mixte,vo:4.5>
 <vo:FF4,vo:norme euro,vo:euro6>
 <vo:FF4,vo:fabriquer par,vo:Ford occasion>
 <vo:FF4,vo:type de véhicule,vo:Berline occasion>
 <vo:FF4,vo:location,vo:Loiret>
 <vo:FF4,vo:price,vo:16750>
\end{lstlisting}

\begin{lstlisting}[captionpos=b, caption=$V_4$ Audi A1 sportback (AA1), label=lst:v4,
	basicstyle=\ttfamily\small,breaklines=true,frame=single]
 <vo:AA1,rdf:type,vo:Automobile>
 <vo:AA1,vo:année,vo:2018-01-01>
 <vo:AA1,vo:mis en circulation,vo:2018-09-15>
 <vo:AA1,vo:contrôle technique,vo:non requis>
 <vo:AA1,vo:kilométrage,vo:20211>
 <vo:AA1,vo:carburant,vo:diesel>
 <vo:AA1,vo:boîte de vistesse,vo:automatique>
 <vo:AA1,vo:coleur extérieure,vo:bleu>
 <vo:AA1,vo:coleur intérieure,vo:noir>
 <vo:AA1,vo:nombre de portes,vo:5>
 <vo:AA1,vo:nombre de places,vo:5>
 <vo:AA1,vo:garranty,vo:12 mois>
 <vo:AA1,vo:puissance fiscale,vo:4>
 <vo:AA1,vo:puissance din,vo:90>
 <vo:AA1,vo:Critique d'Air,vo:2>
 <vo:AA1,vo:emission de CO2,vo:101>
 <vo:AA1,vo:consommation mixte,vo:3.6>
 <vo:AA1,vo:norme euro,vo:euro6>
 <vo:AA1,vo:fabriquer par,vo:Audi occasion>
 <vo:AA1,vo:type de véhicule,vo:Citadine occasion>
 <vo:AA1,vo:location,vo:Yvelines>
 <vo:AA1,vo:price,vo:23200>
\end{lstlisting}	

\begin{lstlisting}[captionpos=b, caption=$V_5$ Citroen C5 aircross (CC5), label=lst:v5,
	basicstyle=\ttfamily\small,breaklines=true,frame=single]
 <vo:CC5,rdf:type,vo:Automobile>
 <vo:CC5,vo:année,vo:2020-01-01>
 <vo:CC5,vo:mis en circulation,vo:2020-06-26>
 <vo:CC5,vo:contrôle technique,vo:non requis>
 <vo:CC5,vo:kilométrage,vo:48368>
 <vo:CC5,vo:carburant,vo:diesel>
 <vo:CC5,vo:boîte de vistesse,vo:mécanique>
 <vo:CC5,vo:coleur extérieure,vo:bleu>
 <vo:CC5,vo:nombre de portes,vo:5>
 <vo:CC5,vo:nombre de places,vo:5>
 <vo:CC5,vo:garranty,vo:12 mois>
 <vo:CC5,vo:puissance fiscale,vo:6>
 <vo:CC5,vo:puissance din,vo:131>
 <vo:CC5,vo:Critique d'Air,vo:2>
 <vo:CC5,vo:emission de CO2,vo:106>
 <vo:CC5,vo:consommation mixte,vo:4.1>
 <vo:CC5,vo:norme euro,vo:euro6>
 <vo:CC5,vo:fabriquer par,vo:Citroen occasion>
 <vo:CC5,vo:type de véhicule,vo:4x4, SUV & Crossover occasion>
 <vo:CC5,vo:location,vo:Yvelines>
 <vo:CC5,vo:price,vo:28890>
\end{lstlisting}
	
\end{document}